\documentclass[%
 aip,
 jmp,%
 amsmath,amssymb,
 reprint,%
]{revtex4-1}

\usepackage{subfigure}
\usepackage{graphicx}
\graphicspath{{./}{./images/}{./images_all/}{./images_all/thermal_energy_delay/}}
\DeclareGraphicsExtensions{.eps}

\usepackage{dcolumn}
\usepackage{bm}

\begin{document}

\preprint{AIP/123-QED}

\title{Energy dissipation and switching delay in stress-induced switching of multiferroic devices in the presence of thermal fluctuations}

\author{Kuntal Roy}
\email{royk@vcu.edu.}
\author{Supriyo Bandyopadhyay}
\affiliation{Department of Electrical and Computer Engineering, Virginia Commonwealth University, Richmond, VA 23284, USA}
\author{Jayasimha Atulasimha}
\affiliation{Department of Mechanical and Nuclear Engineering, Virginia Commonwealth University, Richmond, VA 23284, USA}

\date{\today}

\begin{abstract}
Switching the magnetization of a shape-anisotropic 2-phase multiferroic nanomagnet with voltage-generated stress is known to dissipate very 
little energy ($<$ 1 aJ for a switching time of $\sim$0.5 ns) at 0 K temperature.
Here, we show by solving the stochastic Landau-Lifshitz-Gilbert equation 
that switching can be carried out with $\sim$100\% probability in less than 1 ns while 
dissipating less than 2 aJ at {\it room temperature}. This makes nanomagnetic logic and memory systems,
predicated on stress-induced magnetic reversal, one of the most energy-efficient computing hardware extant.
We also study the dependence of energy dissipation, 
switching delay, and the critical stress needed to switch, on the rate at which stress is ramped up
or down. 
\end{abstract}

\maketitle

\section{\label{sec:introduction}Introduction}

Shape-anisotropic multiferroic nanomagnets, consisting of magnetostrictive layers elastically coupled with piezoelectric layers~\cite{Refworks:164,Refworks:165}, have emerged as attractive storage and switching elements for non-volatile memory and logic systems since they 
are potentially very energy-efficient. Their magnetizations can be switched in less than 1 nanosecond with  energy dissipation 
less than 1 aJ, when no thermal noise is present~\cite{roy11,roy11_2}. This has led to multiple logic proposals incorporating these systems~\cite{RefWorks:154, fasha11, dsouza11}. The magnetization of
the magnet has two (mutually anti-parallel) stable states along the easy axis that encode the
binary bits 0 and 1. The magnetization
is flipped from one stable state to the other by applying a tiny voltage of few tens of millivolts across the piezoelectric layer while constraining the 
magnetostrictive layer from expanding or contracting along its in-plane hard-axis. The voltage generates a strain in the piezoelectric layer, which is then transferred to the magnetostrictive layer. That produces a uniaxial stress in the magnetostrictive layer along its easy-axis and rotates the magnetization towards the in-plane hard axis
as long as the product of the stress and the magnetostrictive coefficient is {\it negative}. Large angle rotations by this method have been demonstrated in recent experiments~\cite{RefWorks:167}, although not in nanoscale.

In this paper, we have studied the switching dynamics of a single-domain magnetostrictive nanomagnet,
subjected to uniaxial stress, in the presence of thermal fluctuations. The dynamics is governed by the stochastic Landau-Lifshitz-Gilbert (LLG) equation~\cite{RefWorks:161,RefWorks:186} that describes the time-evolution of the magnetization vector's orientation 
under various torques. There are {\it three} torques to consider here: torque due to shape anisotropy, torque due to stress, and the torque associated with random thermal fluctuations. With realistic ramp rates (rate at which stress on the magnet is ramped
up or down) a magnet can be switched with $\sim$100\% probability 
with a (thermally averaged) switching delay of $\sim$0.5 ns and (thermally averaged) energy dissipation $\sim$200 $kT$ at room-temperature.
This is very promising for ``beyond-Moore's law'' ultra-low-energy computing~\cite{RefWorks:7,RefWorks:443,RefWorks:123}. Our simulation results show the following: (1) a fast ramp and a sufficiently high stress are required to switch the magnet with high probability in the presence of 
thermal noise, (2) the  stress needed to switch with a given probability increases with decreasing ramp rate, (3) if the ramp rate is too slow, then the switching probability may never approach 100\%
no matter how much stress is applied, (4) the switching probability  
increases monotonically with stress and saturates at $\sim$100\% when the ramp is fast, but exhibits a non-monotonic dependence on stress when the ramp is slow, and (5) the thermal averages of the switching delay and energy dissipation are nearly independent of the ramp rate  if we always switch with the critical stress, which is the minimum value of stress needed to switch with non-zero probability in the
presence of noise.

\begin{figure}
\centering
\includegraphics[width=2.4in]{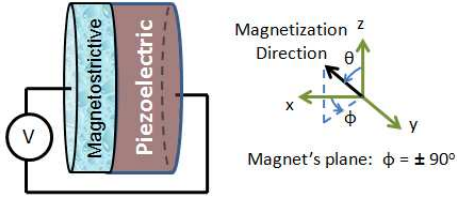}
\caption{\label{fig:multiferroic} A two-phase multiferroic nanomagnet in the shape of 
an elliptical cylinder is stressed with an applied voltage via the $d_{31}$ coupling in the piezoelectric. Mechanical constraints (not
shown) prevent expansion and contraction of the multiferroic along the in-plane hard axis (y-axis).}
\end{figure}

\section{\label{sec:model}Model}

\subsection{Magnetization dynamics of a magnetostrictive nanomagnet in the presence of thermal noise: Solution of the stochastic Landau-Lifshitz-Gilbert equation}

Consider an isolated nanomagnet in the shape of an elliptical cylinder whose elliptical cross section lies in the $y$-$z$ plane with its 
major axis aligned along the z-direction and minor axis along the y-direction (Fig.~\ref{fig:multiferroic}.) The dimension of the major axis is $a$, that of the minor axis is $b$, and the thickness is $l$.  The $z$-axis is the easy axis, the $y$-axis is the in-plane hard axis
and the $x$-axis is the out-of-plane hard axis. Since $l \ll b$, the out-of-plane hard axis is much 
harder than the in-plane hard axis. Let $\theta(t)$ be the polar angle and $\phi(t)$ the azimuthal angle of the 
magnetization vector. 

The total energy of the single-domain, magnetostrictive, polycrystalline nanomagnet, subjected to uniaxial stress along the easy axis (major axis of the ellipse) is the sum of the uniaxial shape anisotropy energy and the uniaxial stress anisotropy energy~\cite{RefWorks:157}. The former is given by~\cite{RefWorks:157} $E_{SHA}(t) = (\mu_0/2) M_s^2 \Omega N_d(t)$, where $M_s$ is the saturation magnetization and $N_d(t)$ is the demagnetization factor expressed 
as~\cite{RefWorks:157} 
\begin{multline}
N_d(t) = N_{d-zz} cos^2\theta(t) + N_{d-yy} sin^2\theta(t) \ sin^2\phi(t) \\ + N_{d-xx} sin^2\theta(t) \, cos^2\phi(t)
\end{multline}
with $N_{d-zz}$, $N_{d-yy}$, and $N_{d-xx}$ being the components of the demagnetization factor along the $z$-axis, $y$-axis, and $x$-axis, respectively~\cite{RefWorks:402}. 
These factors depend on the dimensions of the magnet (values of $a$, $b$ and $l$). 
We choose these dimensions as $a$ = 100 nm, $b$ = 90 nm and $l$ = 6 nm,
which ensures that the magnet has a single domain \cite{RefWorks:133}. These dimensions also determine the shape
anisotropy energy barriers. The in-plane barrier $E_b$, which is the difference between the shape anisotropy
energies when $\theta = 90^{\circ}$ and $\theta = 0^{\circ}, 180^{\circ}$ 
($\phi = \pm 90^{\circ}$) determines the static error probability, which is the probability 
of spontaneous magnetization reversal due to thermal noise. This probability is $\exp \left [
-E_b/kT \right ]$. For the dimensions and material chosen, $E_b$ = 44 $kT$ at room temperature, so that the static error probability at room temperature is
$e^{-44}$. 

The stress anisotropy energy is given by~\cite{RefWorks:157} $E_{STA}(t) = - (3/2) \lambda_s \sigma(t) \Omega \, cos^2\theta(t)$, where $(3/2) \lambda_s$ is the magnetostriction coefficient of the nanomagnet and $\sigma(t)$ is the stress 
at an instant of time $t$. Note that a positive $\lambda_s \sigma(t)$ product will favor alignment of the magnetization along 
the major axis ($z$-axis), while a negative $\lambda_s \sigma(t)$ product will favor alignment along the minor axis ($y$-axis), because that will minimize $E_{STA}(t)$. In our convention, a compressive stress is negative and tensile stress is positive. Therefore, in a material 
like Terfenol-D that has positive $\lambda_s$, a compressive stress will favor alignment along the minor axis (in-plane hard axis), and tensile along the major axis (easy axis)~\cite{roy11}. 

At any instant of time $t$, the total energy of the nanomagnet can be expressed as 
\begin{equation}
E(t) = E(\theta(t),\phi(t),\sigma(t)) = B(t) sin^2\theta(t) + C(t)
\end{equation}
where 
\begin{subequations}
\begin{align}
B(t) &= B_0(t) + B_{stress}(t) \displaybreak[3]\\
B_0(t) &= (\mu_0/2) \, M_s^2 \Omega \lbrack N_{d-xx} cos^2\phi(t) \nonumber \\&+ N_{d-yy} sin^2\phi(t) - N_{d-zz}\rbrack \displaybreak[3]\\
B_{stress}(t) &= (3/2) \lambda_s \sigma(t) \Omega \displaybreak[3]\\
C(t) &= (\mu_0/2) M_s^2 \Omega N_{d-zz} - (3/2) \lambda_s \sigma(t) \Omega. \displaybreak[3]
\end{align}
\end{subequations}

The torque acting on the magnetization per unit volume due to shape and stress anisotropy is
\begin{eqnarray}
\mathbf{T_E} (t) &=& - \mathbf{n_m}(t) \times \nabla E(\theta(t),\phi(t),\sigma(t)) \nonumber\\
&=& - 2 B(t) sin\theta(t) cos\theta(t) \,\mathbf{\hat{e}_\phi}  - B_{0e}(t) sin\theta (t) \,\mathbf{\hat{e}_\theta}, 							 
\label{eq:stress_torque}
\end{eqnarray}
\noindent
where $B_{0e}(t) = (\mu_0/2) \, M_s^2 \Omega (N_{d-xx}-N_{d-yy}) sin(2\phi(t))$. 

The torque due to thermal fluctuations is treated via a random magnetic field $\mathbf{h}(t)$ and is expressed as 
\begin{equation}
\mathbf{h}(t) = h_x(t)\mathbf{\hat{e}_x} + h_y(t)\mathbf{\hat{e}_y} + h_z(t)\mathbf{\hat{e}_z}
\end{equation}
\noindent
where $h_x(t)$, $h_y(t)$, and $h_z(t)$ are the three components of the random thermal field $\mathbf{h}(t)$ in $x$-, $y$-, and $z$-direction, respectively 
in Cartesian coordinates. We assume the properties of the random field $\mathbf{h}(t)$ as described in Ref.~[\onlinecite{RefWorks:186}].
Accordingly, the random thermal field can be expressed as
\begin{equation}
h_i(t) = \sqrt{\frac{2 \alpha kT}{|\gamma| (1+\alpha^2) M_V \Delta t}} \; G_{(0,1)}(t) \qquad (i=x,y,z)
\label{eq:ht}
\end{equation}
\noindent
where $1/\Delta t$ is proportional to the attempt frequency of the thermal field. Consequently,
$\Delta t$ should be the simulation time-step used to simulate switching trajectories in the 
presence of random thermal torque. The quantity $G_{(0,1)}(t)$ is a Gaussian distribution with zero mean and unit variance~\cite{RefWorks:388}. 

The thermal torque can be written as
\begin{equation}
\mathbf{T_{TH}}(t) = M_V\,\mathbf{n_m}(t) \times \mathbf{h}(t) = P_\theta(t)\,\mathbf{\hat{e}_\phi} - P_\phi(t)\,\mathbf{\hat{e}_\theta}
\end{equation}
\noindent
where
\begin{eqnarray}
P_\theta(t) &=& M_V\lbrack h_x(t)\,cos\theta(t)\,cos\phi(t) + h_y(t)\,cos\theta(t)sin\phi(t) \nonumber\\ && - h_z(t)\,sin\theta(t) \rbrack\\
P_\phi(t) &=& M_V \lbrack h_y(t)\,cos\phi(t) -h_x(t)\,sin\phi(t)\rbrack.
\label{eq:thermal_parts}
\end{eqnarray}
\noindent

The magnetization dynamics under the action of the torques $\mathbf{T_{E}}(t)$ and 
$\mathbf{T_{TH}}(t)$ is described by the stochastic Landau-Lifshitz-Gilbert (LLG) equation as follows.
\begin{equation}
\cfrac{d\mathbf{n_m}(t)}{dt} - \alpha \left(\mathbf{n_m}(t) \times \cfrac{d\mathbf{n_m}(t)}
{dt} \right) = -\cfrac{|\gamma|}{M_V} \left\lbrack \mathbf{T_E}(t) +  \mathbf{T_{TH}}(t)\right\rbrack
\label{LLG}
\end{equation}
where $\alpha$ is the dimensionless phenomenological Gilbert damping constant, $\gamma = 2\mu_B \mu_0/\hbar$ 
is the gyromagnetic ratio for electrons and is equal to $2.21\times 10^5$ (rad.m).(A.s)$^{-1}$, $\mu_B$ is the Bohr magneton, 
and $M_V= \mu_0 M_s \Omega$. 

From the last equation, we get the following coupled equations for the dynamics of $\theta(t)$ and $\phi(t)$.
\begin{multline}
\left(1+\alpha^2 \right) \cfrac{d\theta(t)}{dt} = \frac{|\gamma|}{M_V} \lbrack 
 B_{0e}(t) sin\theta(t) \\ - 2\alpha B(t) sin\theta (t)cos\theta (t) + \left(\alpha P_\theta + P_\phi \right) \rbrack.
 \label{eq:theta_dynamics}
\end{multline}
\begin{multline}
\left(1+\alpha^2 \right) \cfrac{d \phi(t)}{dt} = \frac{|\gamma|}{M_V} 
\lbrack \alpha B_{0e}(t) + 2 B(t) cos\theta(t) \\ - {sin^{-1}\theta(t)} \left(P_\theta - \alpha P_\phi \right) \rbrack. \quad
	(sin\theta \neq 0.)
\label{eq:phi_dynamics}
\end{multline}
These equations describe the magnetization dynamics, namely the temporal evolution of the magnetization
vector's orientation, in the presence of thermal noise.

\subsection{\label{sec:initial_thermal} Fluctuation of magnetization around the easy
axis (stable orientation) due to thermal noise}

 The torque on the magnetization vector due to shape and stress anisotropy vanishes when $sin \theta = 0$ [see Equation (\ref{eq:stress_torque})], i.e. when the magnetization 
vector is aligned along the easy axis. That is
why $\theta = 0^{\circ}, 180^{\circ}$ are called {\it stagnation points}.
Only thermal fluctuations can budge the magnetization vector from 
the easy axis. To see this, consider the 
situation when $\theta=180^\circ$. We get:
\begin{equation}
\phi(t) = tan^{-1} \left( \frac{\alpha h_y(t) + h_x(t)}{h_y(t) - \alpha h_x(t)} \right),
\label{eq:phi_t_thermal}
\end{equation}
\begin{equation}
\theta'(t) = -|\gamma| \frac{h_x^2(t) + h_y^2(t)}{\sqrt{(h_y(t)-\alpha h_x(t))^2 + (\alpha h_y(t) + h_x(t))^2}}.
\end{equation}
\noindent
We can see clearly from the above equation that thermal torque can deflect the magnetization 
from the easy axis since the time rate of change of $\theta(t)$ [i.e., $\theta'(t)$ is non-zero
in the presence of the thermal field. Note that the initial deflection from the easy axis due to the thermal torque does not depend on the component of the random thermal field along the $z$-axis, i.e., $h_z(t)$, which is a consequence of having $\pm$$z$-axis as the easy axes of the nanomagnet. However, once the magnetization direction is even slightly deflected from the easy axis, all  three components of the random thermal field along the $x$-, $y$-, and $z$-direction would come into play and affect the deflection. 

\subsection{\label{sec:initial_thermal_mag_field} Thermal distribution of the
initial orientation of the magnetization vector}

The thermal distributions of $\theta$ and $\phi$ in the unstressed magnet are found 
by solving the Equations (\ref{eq:theta_dynamics}) and (\ref{eq:phi_dynamics}) while setting $B_{stress}$ = 0.
This will yield the distribution of the magnetization vector's initial orientation when stress is turned on.
The $\theta$-distribution is Boltzmann peaked at $\theta$ = 0$^{\circ}$ or 180$^{\circ}$, while the $\phi$-distribution is Gaussian peaked at $\phi = \pm 90^{\circ}$ (Ref.~[\onlinecite{roy11_5}]). Since the most probable value of $\theta$ is either 0$^{\circ}$
or 180$^{\circ}$, where 
stress is ineffective ({\it stagnation point}), there are long
tails in the switching delay distribution. They are due to the fact that
when we start out from $\theta = 0^{\circ}, 180^{\circ}$, we have to
wait a while before thermal kick sets the switching in motion. Thus, switching trajectories initiating
from a stagnation point are very slow~\cite{RefWorks:198,RefWorks:480}. 

In order to eliminate the long tails in the switching delay distribution and thus decrease the mean switching 
delay, one can apply a small static bias magnetic field that will shift the peak of $\theta_{initial}$ distribution away from the easy axis, so that the most probable starting orientation will no
longer be a stagnation point. This field is applied along the out-of-plane hard axis (+$x$-direction) 
so that the potential energy due to the applied magnetic field becomes $E_{mag}(t) = - M_V H\, sin\theta(t)\,cos\phi(t)$, where
$H$ is the magnitude of magnetic field. The torque generated due to this field is $\mathbf{T_M} (t) = - \mathbf{n_m}(t) \times \nabla E_{mag}(\theta(t),\phi(t))$. The presence of this field will modify Equations~\eqref{eq:theta_dynamics} and~\eqref{eq:phi_dynamics}  to 
\begin{multline}
\left(1+\alpha^2 \right) \cfrac{d\theta(t)}{dt} = \frac{|\gamma|}{M_V} \lbrack 
 B_{0e}(t) sin\theta(t) - 2\alpha B(t) sin\theta (t)cos\theta (t) 
 \\  + \alpha M_V H\, cos\theta(t)\,cos\phi(t) - M_V H\, sin\phi(t) 
 \\ + \left(\alpha P_\theta + P_\phi \right) \rbrack,
 \label{eq:theta_dynamics_mag}
\end{multline}
\begin{multline}
\left(1+\alpha^2 \right) \cfrac{d \phi(t)}{dt} = \frac{|\gamma|}{M_V} 
\lbrack \alpha B_{0e}(t) + 2 B(t) cos\theta(t) 
\\ - {sin^{-1}\theta(t)} \left(M_V H\, cos\theta(t)\,cos\phi(t) + \alpha M_V H\, sin\phi(t)   \right)
\\ - {sin^{-1}\theta(t)} \left(P_\theta - \alpha P_\phi \right) \rbrack. \quad	(sin\theta \neq 0.)
\label{eq:phi_dynamics_mag}
\end{multline}
\noindent

The bias field also makes the potential energy profile of the magnet asymmetric in $\phi$-space and the energy minimum 
 will be shifted from $\phi_{min}=\pm90^\circ$ (the plane of the magnet) to
\begin{equation}
\phi_{min} = cos^{-1}\left\lbrack \frac{H}{M_s (N_{d-xx} - N_{d-yy})} \right\rbrack.
\end{equation}
\noindent
However, the profile will remain symmetric in $\theta$-space, with $\theta=0^\circ$ and $\theta=180^\circ$ remaining as the minimum energy locations. With
 the parameters used in this paper, a bias magnetic field of flux density 40 mT would make $\phi_{min} \simeq 87^\circ$. Application of the bias magnetic field will also reduce the 
in-plane shape anisotropy energy barrier from 44 $kT$ to 36 $kT$ at room temperature. We assume that a permanent magnet will  be employed to produce the bias field and thus 
will not affect the energy dissipated during switching.

\subsection{\label{sec:enenrgy_dissipation} Energy Dissipation}

The energy dissipated during switching has two components: (1) the energy dissipated in the switching circuit that applies the stress on the nanomagnet by generating a voltage, and (2) the energy dissipated internally in the nanomagnet because of Gilbert damping. We will term the first component `$CV^2$' dissipation, where $C$ and $V$ denote the capacitance of the piezoelectric layer and the applied voltage, respectively. If the voltage is turned on or off abruptly, i.e. the ramp rate is infinite, 
then the energy dissipated during either turn on or turn off is $(1/2)CV^2$.
However, if the ramp rate is finite, then this energy is reduced and its exact value
will depend on the ramp
duration or ramp rate. We calculate it following the same procedure described in Ref.~[\onlinecite{roy11_2}]. The second component, which is the internal energy dissipation $E_d$, is given by the expression $\int_0^{\tau}P_d(t) dt$, where $\tau$ is the switching delay and $P_d(t)$ is the power dissipated during switching \cite{RefWorks:319,RefWorks:124}
\begin{equation}
P_d(t) = \frac{\alpha \, |\gamma|}{(1+\alpha^2) M_V} \, \left| \mathbf{T_E}(t) + \mathbf{T_M}(t)\right|^2 .
\label{eq:power_dissipation}
\end{equation}
\noindent
 We sum up the power $P_d(t)$ dissipated during the entire switching period to get the corresponding energy dissipation $E_d$ and add that to the 
`$CV^2$' dissipation in the switching circuit to find the total dissipation $E_{total}$. The average power dissipated during switching is simply $E_d/\tau$.

There is no net dissipation due to random thermal torque since the mean of the random thermal field is zero. However, that does not mean that the temperature has no effect on either $E_d$ or the `$CV^2$'
dissipation. It affects $E_d$ since it raises the critical stress needed to switch with non-zero 
probability and it also affects the stress needed to switch with a given probability. Furthermore,
it affects `$CV^2$' because $V$ must exceed the thermal noise voltage \cite{RefWorks:549} to prevent random 
switching due to noise. In other words, we must enforce $CV^2 > kT$. For the estimated capacitance
of our structure (2.6 fF), this translates to $V >$ 1.3 mV.

\begin{figure}
\centering
\subfigure[]{\label{fig:thermal_theta_distribution_terfenolD_mag_field}\includegraphics[width=2.8in]
{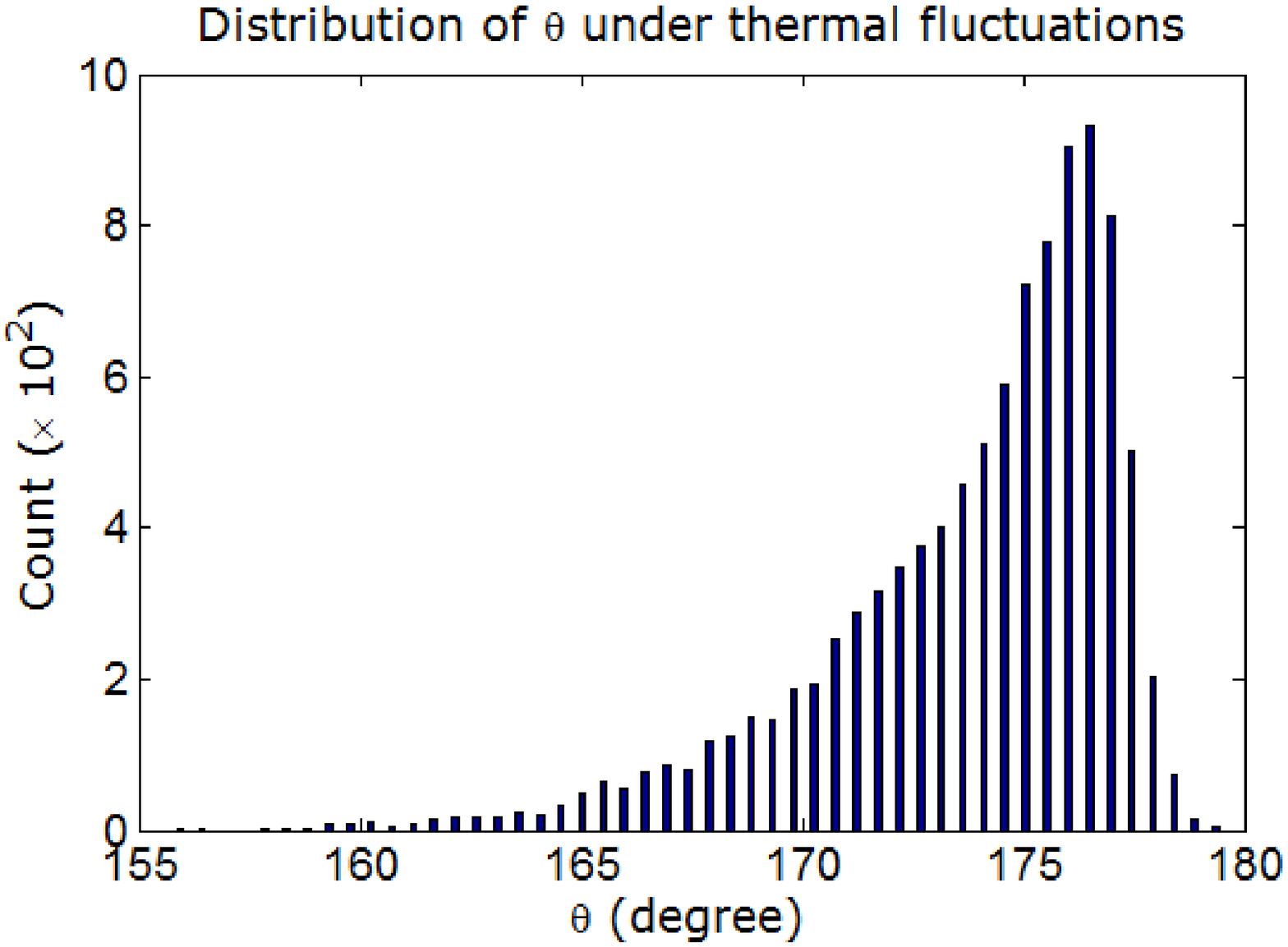}}
\subfigure[]{\label{fig:thermal_phi_distribution_terfenolD_mag_field}\includegraphics[width=2.8in]
{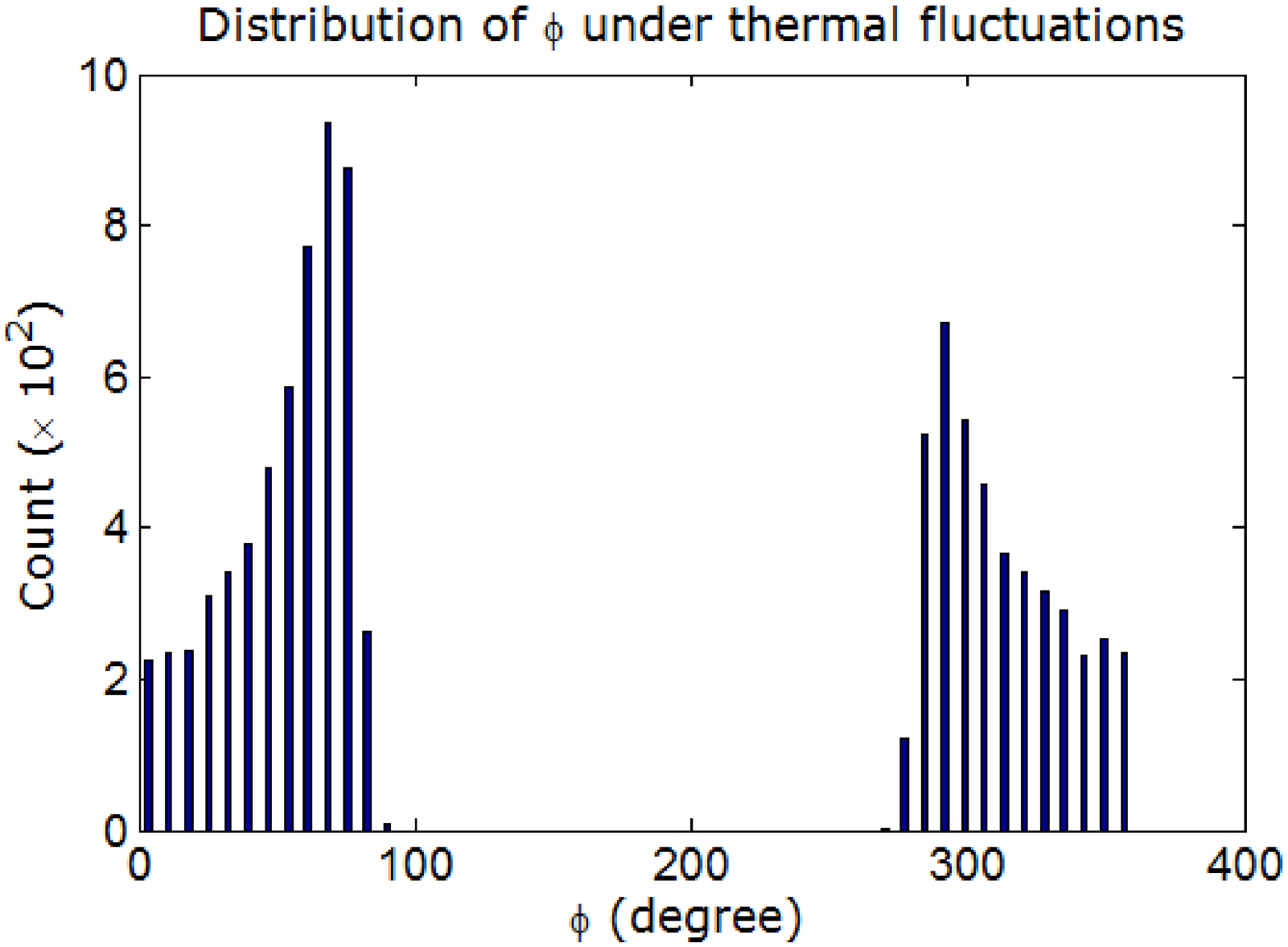}}
\caption{\label{fig:thermal_theta_phi_distribution_terfenolD_mag_field} Distribution of polar angle $\theta_{initial}$ and azimuthal 
angle $\phi_{initial}$ due to thermal fluctuations at room temperature (300 K) when a magnetic field of
flux density 40 mT is applied along the out-of-plane hard axis (+$x$-direction).
(a) Distribution of polar angle $\theta_{initial}$ at room temperature (300 K). The mean of the distribution is $173.7^\circ$, and
 the most likely value is 175.8$^{\circ}$.
(b) Distribution of the azimuthal angle $\phi_{initial}$ due to thermal fluctuations at room temperature (300 K). There are two distributions with  peaks centered at $\sim$65$^\circ$ and $\sim$295$^\circ$.}
\end{figure}

\section{\label{sec:results}Simulation results and Discussions}

In our simulations, we consider the magnetostrictive layer to be  made of polycrystalline Terfenol-D that has the following material properties -- Young's modulus (Y): 8$\times$10$^{10}$ Pa, magnetostrictive coefficient ($(3/2)\lambda_s$): +90$\times$10$^{-5}$, saturation magnetization ($M_s$):  8$\times$10$^5$ A/m, and Gilbert's damping constant ($\alpha$): 0.1 (Refs.~[\onlinecite{RefWorks:179,RefWorks:176,RefWorks:178, materials}]). For the piezoelectric layer, we use lead-zirconate-titanate (PZT), which has a dielectric constant of 1000. The PZT layer is assumed to be four times thicker than the magnetostrictive layer so that any strain generated in it is transferred almost completely to the magnetostrictive layer~\cite{roy11}. We will assume that the maximum strain that can be generated in the PZT layer is 500 ppm \cite{RefWorks:170}, which would require a voltage of 111 mV because $d_{31}$=1.8$\times$10$^{-10}$ m/V for PZT \cite{pzt2}. The corresponding stress is the product of the generated strain ($500\times10^{-6}$) and the Young's modulus of the  magnetostrictive layer. Hence, 40 MPa is the maximum magnitude of stress that can be generated on the nanomagnet.

We assume that when stress is applied to initiate switching, the magnetization vector
starts out from near the south pole ($\theta \simeq 180^\circ$) with a certain ($\theta_{initial}$,$\phi_{initial}$) picked from the initial angle distributions at the given
temperature. Stress is ramped up linearly and kept constant until the magnetization reaches the plane defined by the in-plane and the out-of-plane hard axis (i.e. the $x$-$y$ plane, $\theta = 90^\circ$). This plane is always reached sooner
or later since the energy minimum of the stressed magnet in $\theta$-space is at $\theta = 90^{\circ}$.
Thermal fluctuations can introduce a spread in the time it takes to reach the 
$x$-$y$ plane but cannot prevent the magnetization from reaching it ultimately if the 
stress is so large that the energy minimum at $\theta = 90^{\circ}$ is more than a few $kT$ deep.

As soon as the magnetization reaches the $x$-$y$ plane, the
 stress is ramped down at the same rate at which it was ramped up, and reversed in magnitude to facilitate switching. 
The magnetization dynamics ensures that $\theta$ continues to rotate towards $0^{\circ}$
with very high probability. When $\theta$ becomes $ \leq 5^\circ$, switching is deemed to have completed. A moderately large number (10,000) of simulations, with their corresponding ($\theta_{initial}$,$\phi_{initial}$) picked from the initial angle distributions, are performed for each value of stress and ramp duration to generate the simulation results in this paper. 

Fig.~\ref{fig:thermal_theta_phi_distribution_terfenolD_mag_field} shows the distributions of initial angles $\theta_{initial}$ and $\phi_{initial}$ in the presence of thermal fluctuations and an applied bias magnetic field along the +$x$-direction. The latter has shifted the peak from the easy axis ($\theta = 180^{\circ}$). In  Fig.~\ref{fig:thermal_phi_distribution_terfenolD_mag_field}, the $\phi_{initial}$ distribution has two peaks and resides mostly within the interval [-90$^\circ$,+90$^\circ$] since the bias magnetic field is applied in the +$x$-direction. Because the magnetization 
vector starts out from near the south pole ($\theta \simeq 180^\circ$) when stress is turned on, the effective torque on the magnetization $ -|\gamma|/\left ( 1 + \alpha^2 \right ) \mathbf{M} \times \mathbf{H}$
due to the +$x$-directed magnetic field is such that the magnetization prefers the $\phi$-quadrant (0$^\circ$,90$^\circ$) slightly over the $\phi$-quadrant (270$^\circ$,360$^\circ$), which is the reason for the asymmetry in the two distributions of $\phi_{initial}$. Consequently, when the magnetization vector starts out from $\theta \simeq 180^{\circ}$, the initial azimuthal angle $\phi_{initial}$ is a little more likely to be in the quadrant (0$^\circ$,90$^\circ$) than the quadrant (270$^\circ$,360$^\circ$).

\begin{figure}
\centering
\includegraphics[width=2.8in]{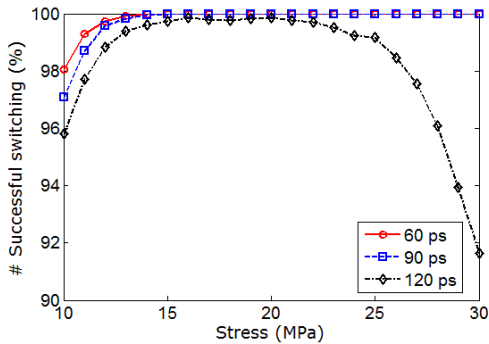}
\caption{\label{fig:thermal_stress_success_ramp_time} Percentage of
successful switching events among the simulated switching trajectories (or the switching probability)
at room temperature in a Terfenol-D/PZT multiferroic nanomagnet 
when subjected to stress between 10 MPa and 30 MPa. The stress at which 
switching becomes $\sim$100\% successful increases with ramp duration. For large ramp duration (120 ps) or slow ramp rate, $\sim$100\% switching probability is unachievable.}
\end{figure}

Fig.~\ref{fig:thermal_stress_success_ramp_time} shows the switching probability as a function of stress for different ramp durations (60 ps, 90 ps, 120 ps) at room temperature (300 K). The minimum stress needed to switch the magnetization with $\sim$100\% probability at 0 K is $\sim$5 MPa, but at 300 K, it increases to $\sim$14 MPa for 60 ps ramp duration and $\sim$17 MPa for 90 ps ramp duration. At low stress levels, the switching probability increases with stress, regardless of the ramp rate. This happens
because a higher stress mitigates the detrimental effects of thermal fluctuations more when magnetization reaches the $x$-$y$ plane and thus conducive to more success rate of switching. This feature is independent of the ramp rate.

Once the magnetization vector crosses the $x$-$y$ plane (i.e. in the second half of
switching), the ramp rate becomes important.
Now, the stress initially applied to cause switching 
becomes harmful and impedes switching. That happens because it causes the energy minimum to be located at $\theta = 90^{\circ}$, which will make the magnetization backtrack towards this location during the 
second half. This is why stress must be
removed or reversed immediately upon crossing the $x$-$y$ plane so that the  energy minimum quickly moves back to $\theta
 = 0^{\circ}, 180^{\circ}$, and the magnetization vector rotates towards
$\theta = 0^{\circ}$. If the removal rate is fast, then the 
success probability remains high since the harmful stress
does not stay active long enough to cause significant backtracking.
However, if the ramp rate is too slow, then significant backtracking occurs whereupon the magnetization
vector returns to the $x$-$y$ plane and 
thermal torques can subsequently kick it to the starting position at $\theta \simeq 180^{\circ}$,
causing switching failure. That is why the switching probability drops with decreasing ramp rate.

The same effect also explains the non-monotonic stress dependence of the
switching probability when the ramp rate is slow. During the first half of the switching,
when $\theta$ is in the quadrant [180$^{\circ}$, 90$^{\circ}$], a higher stress is helpful
since it provides a larger torque to move towards the $x$-$y$ plane, but during the second half,
when $\theta$ is in the quadrant [90$^{\circ}$, 0$^{\circ}$],
a higher stress is harmful since it increases the chance of backtracking. These two counteracting effects 
are the reason for the non-monotonic dependence of the success probability on stress in the case of the slowest ramp rate.

\begin{figure}
\centering
\includegraphics[width=2.8in]{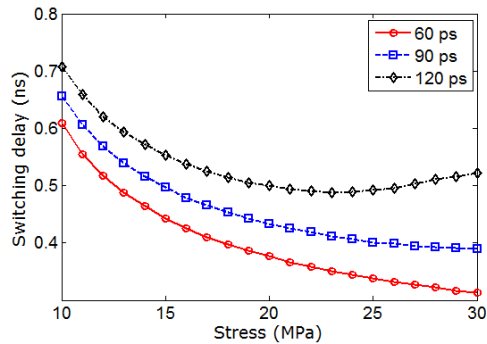}
\caption{\label{fig:thermal_mean_stress_delay} The thermal mean of the switching delay (at 300 K) versus stress (10-30 MPa) for different ramp durations (60 ps, 90 ps, 120 ps). Switching may fail at low stress levels and also at high stress levels for long ramp durations. Failed attempts are excluded when computing the mean.}
\end{figure}

\begin{figure}
\centering
\includegraphics[width=2.8in]{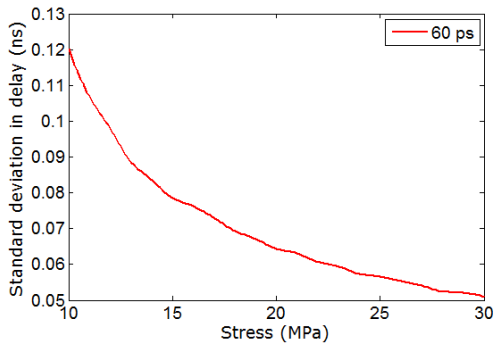}
\caption{\label{fig:thermal_stress_delay_std} The standard deviations in switching delay versus stress (10-30 MPa) for 60 ps ramp duration at 300 K. We consider only the successful switching events in determining the standard deviations. The standard deviations in switching delay  for other ramp durations are of similar magnitudes and show similar trends.}
\end{figure}

Fig.~\ref{fig:thermal_mean_stress_delay} shows the thermally averaged switching delay versus stress for different ramp durations. Only successful switching events are counted here
since the switching delay will be infinity for an unsuccessful event. For a given stress, decreasing the ramp duration (or increasing the ramp rate) decreases the switching delay because the stress reaches its maximum value quicker and hence switches the magnetization faster. For ramp durations of 60 ps and 90 ps, the switching delay decreases with increasing stress since the torque, which rotates the magnetization, increases when stress increases. However, for 120 ps ramp duration, the dependence is non-monotonic, because of the same reasons that caused the non-monotonicity in Fig. 3. Too high a stress is harmful
during the second half of the switching since it increases the chances of backtracking. Even if backtracking can be 
overcome and successful switching ultimately takes place, temporary backtracking still
increases the switching delay.

Fig.~\ref{fig:thermal_stress_delay_std} shows the standard deviation in switching delay versus stress for 60 ps ramp duration. At higher values of stress, the torque due to stress dominates over the random thermal torque that causes the spread in the switching delay. That makes the distribution more peaked as we 
increase the stress.

\begin{figure}
\centering
\includegraphics[width=2.8in]{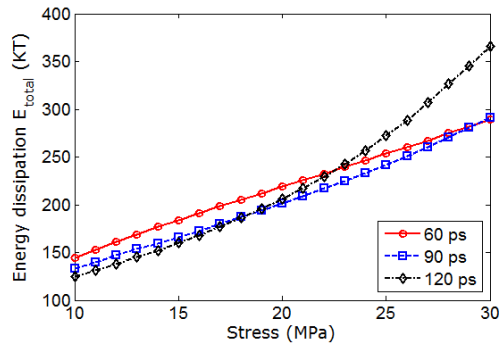}
\caption{\label{fig:thermal_mean_stress_energy} Thermal mean of the total energy dissipation versus stress (10-30 MPa) for different ramp durations (60 ps, 90 ps, 120 ps). Once again, failed switching attempts are excluded when computing the mean.}
\end{figure}

\begin{figure}
\centering
\includegraphics[width=2.8in]{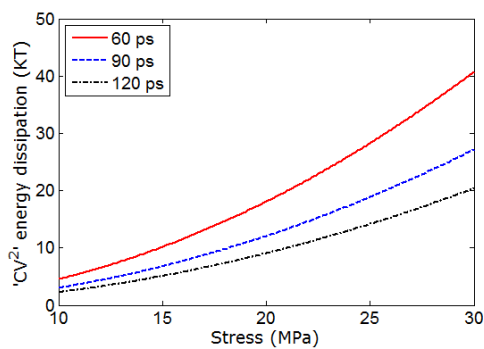}
\caption{\label{fig:thermal_stress_energy_CV2} The `$CV^2$' energy dissipation in the external circuit as a function of stress for different ramp durations.}
\end{figure}

Fig.~\ref{fig:thermal_mean_stress_energy} shows the thermal mean of the total energy dissipated to switch the magnetization as a function of stress for 
different ramp durations. The average power dissipation ($E_{total}/\tau$) increases with stress for a given ramp duration and decreases with increasing ramp duration for a given stress. More stress 
requires more `$CV^2$' dissipation and also more internal dissipation because it results in a higher torque. Slower switching decreases the power dissipation since it makes the switching more
adiabatic. However, since the switching delay curves show the opposite trend (see Fig.~\ref{fig:thermal_mean_stress_delay}), the energy dissipation curves in  Fig.~\ref{fig:thermal_mean_stress_energy} exhibit the cross-overs. 

Fig.~\ref{fig:thermal_stress_energy_CV2} shows the `$CV^2$' energy dissipation in the switching circuitry versus stress. Increasing stress 
requires increasing the voltage $V$, which is why the `$CV^2$' energy dissipation increases rapidly with stress. This dissipation however is a small fraction of the total energy dissipation ($<$ 15\%) since a very small voltage is required to switch the magnetization of a multiferroic nanomagnet with stress. The `$CV^2$' dissipation decreases when the ramp duration increases because then the switching becomes more `adiabatic' and hence less dissipative. This component of the energy dissipation would have been 
several orders of magnitude higher had we switched the magnetization with an external magnetic field~\cite{RefWorks:142} or spin-transfer torque~\cite{RefWorks:7}.

\begin{figure}
\centering
\subfigure[]{\label{fig:thermal_distribution_delay_60ps_15MPa_mag}\includegraphics[width=2.8in]
{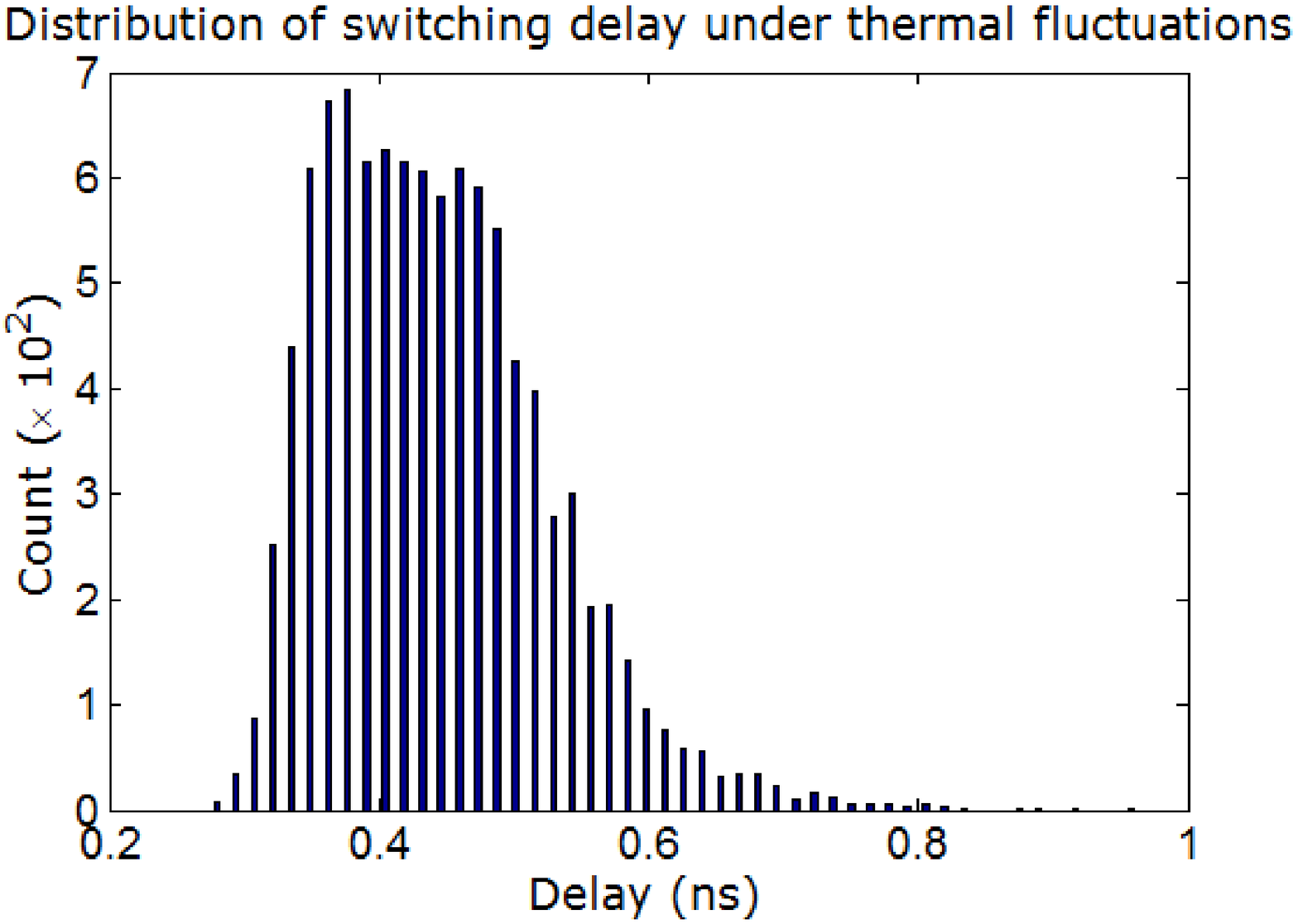}}
\subfigure[]{\label{fig:thermal_distribution_energy_60ps_15MPa_mag}\includegraphics[width=2.8in]
{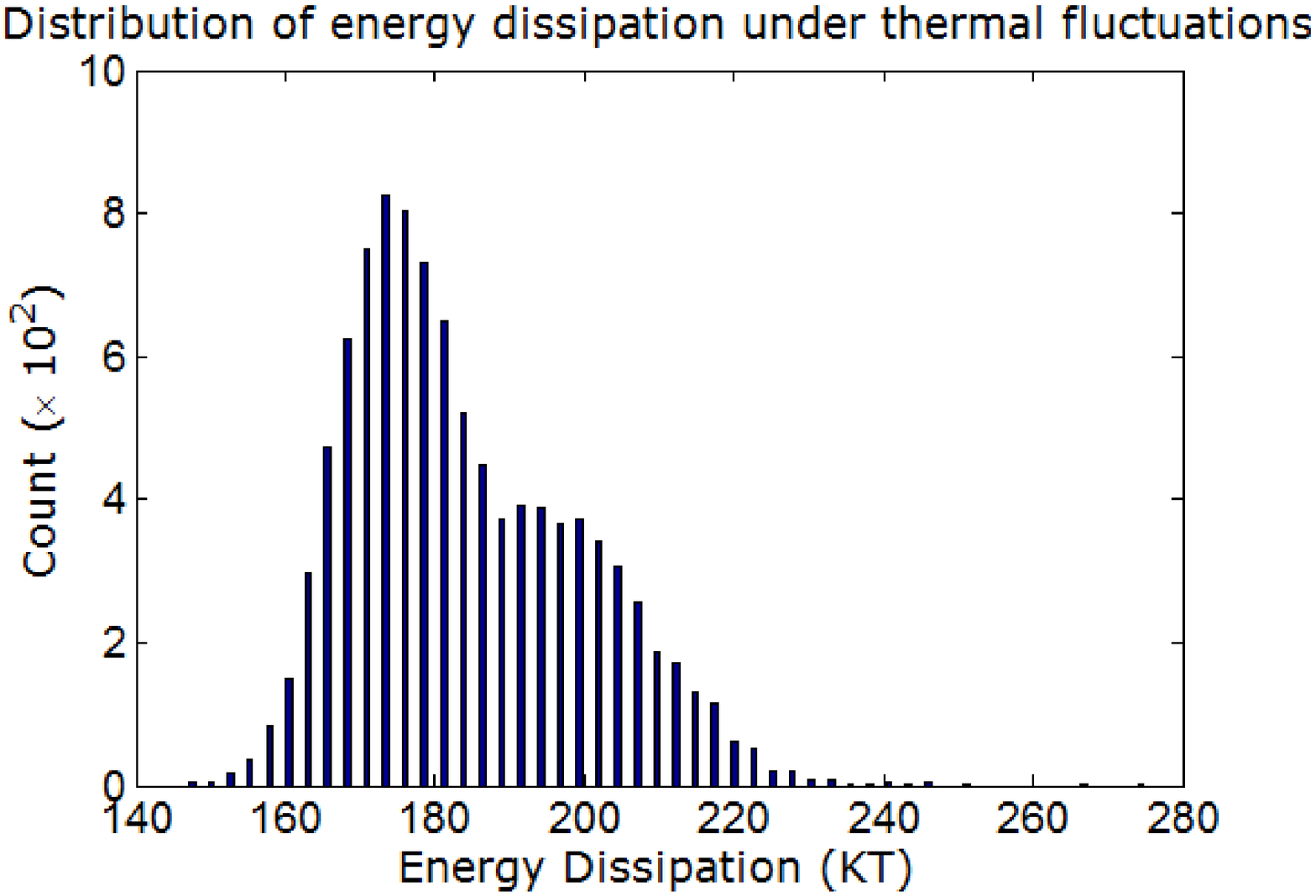}}
\caption{\label{fig:thermal_distribution_60ps_15MPa_mag} Delay and energy distributions for 15 MPa applied stress and 60 ps ramp duration 
at room temperature (300 K). (a) Distribution of the switching delay. The mean and standard deviation of the distribution are 0.44 ns and 83 ps, respectively. (d) Distribution of energy dissipation. The mean and standard deviation of the distribution are 184 $kT$ and 15.5 $kT$ at room temperature, respectively.
}
\end{figure}

Fig.~\ref{fig:thermal_distribution_60ps_15MPa_mag} shows the delay and energy distributions in the presence of room-temperature thermal fluctuations for 15 MPa stress and 60 ps ramp duration. The high-delay tail in  Fig.~\ref{fig:thermal_distribution_delay_60ps_15MPa_mag} is associated 
with those switching trajectories that start very close to $\theta = 180^\circ$ which is a stagnation
point. In such trajectories, the starting torque is vanishingly small, which makes the switching 
sluggish at the beginning. During this time, switching also becomes susceptible to backtracking 
because of thermal fluctuations, which increases the delay further. Since the energy dissipation is
the product of the mean power dissipation and the switching delay, similar behavior is found
in Fig.~\ref{fig:thermal_distribution_energy_60ps_15MPa_mag}.

\begin{figure}
\centering
\subfigure[]{\label{fig:dynamics_success_60ps_10MPa_mag}\includegraphics[width=2.8in]
{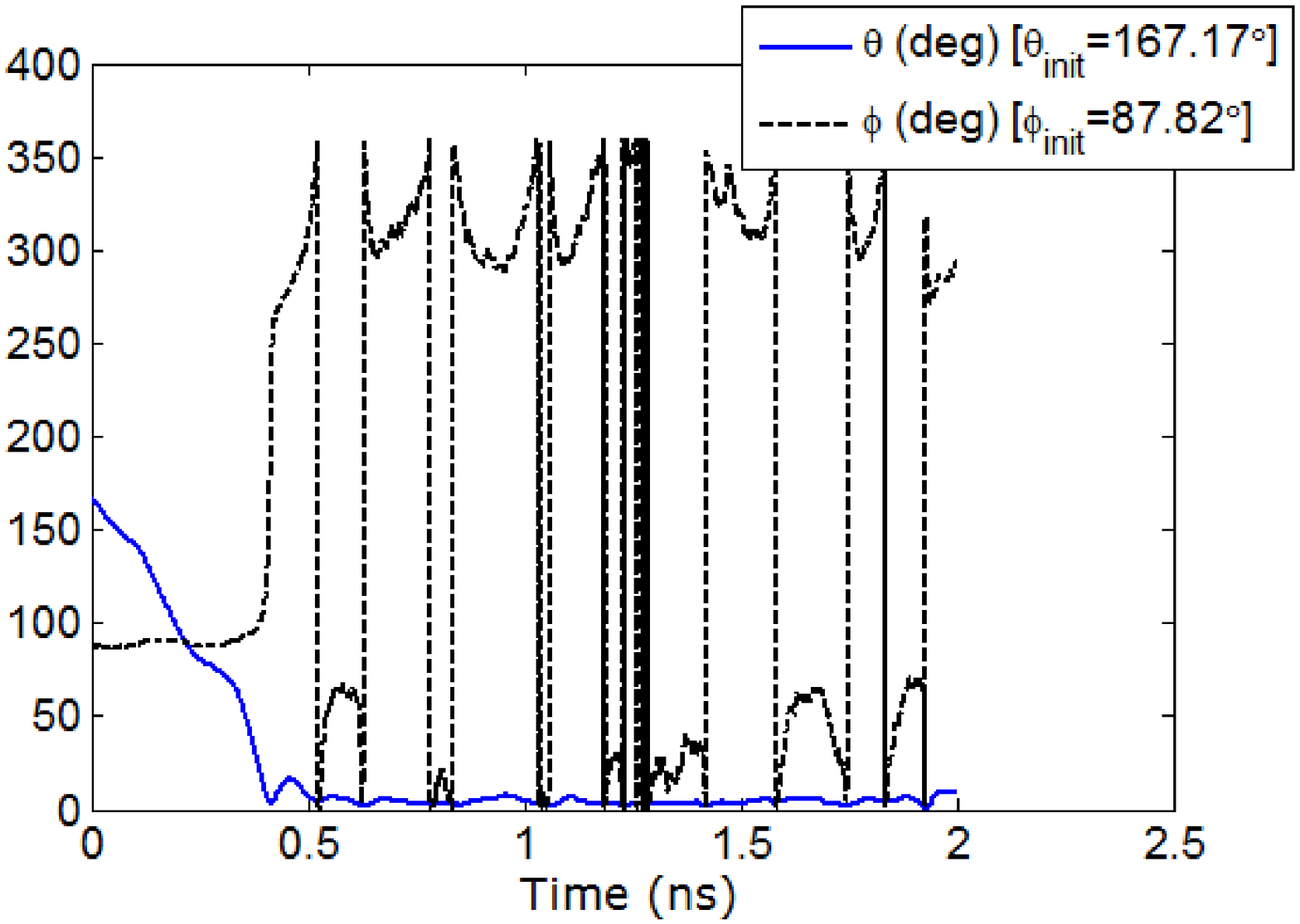}}
\subfigure[]{\label{fig:dynamics_failed_60ps_10MPa_mag}\includegraphics[width=2.8in]
{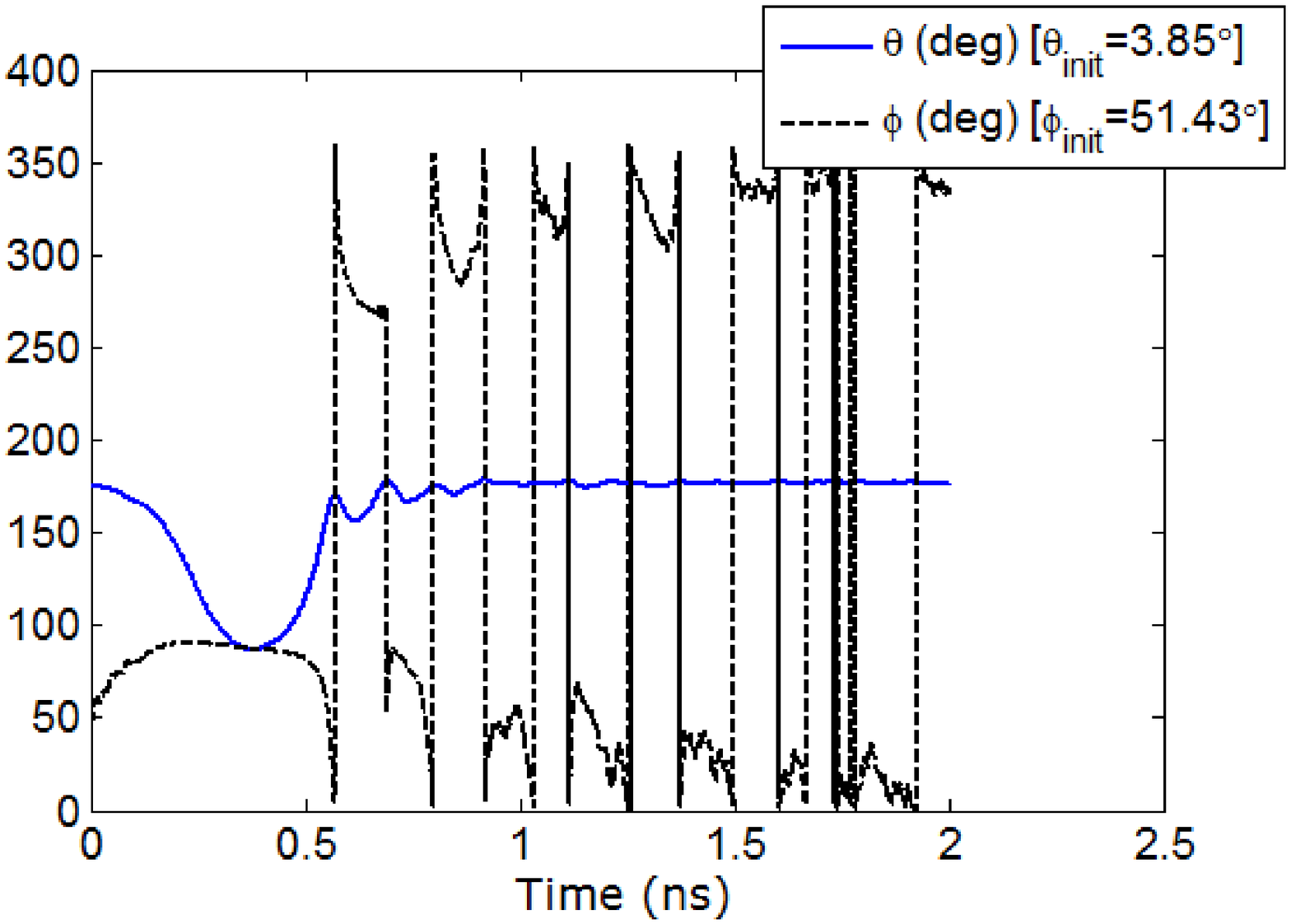}}
\caption{\label{fig:dynamics_examples} Temporal evolution of the polar 
angle $\theta(t)$ and azimuthal angle $\phi(t)$ for 10 MPa applied stress and 60 ps ramp duration. Simulations are carried out for room temperature (300 K). (a) Magnetization switches successfully. (b) Magnetization fails to switch and backtracks towards the initial state.}
\end{figure}

Fig.~\ref{fig:dynamics_examples} shows two examples of switching dynamics when the applied
stress is 10 MPa  and the ramp duration is 60 ps. In Fig.~\ref{fig:dynamics_success_60ps_10MPa_mag}, magnetization switches successfully. Thermal fluctuations cause the ripples because of temporary backtracking
but $\theta$ switches from $\sim$180$^{\circ}$ to $\sim$0$^{\circ}$ finally.
Note that despite appearances, $\phi$ is not changing discretely. When it crosses 360$^{\circ}$,
it re-enters the quadrant [$0^{\circ}, 90^{\circ}$], which is why it appears as if there is 
a discrete jump in the value of $\phi$ in Fig.~\ref{fig:dynamics_examples}.
On the other hand, Fig.~\ref{fig:dynamics_failed_60ps_10MPa_mag} shows a failed switching dynamic. Here, the magnetization backtracks towards $\theta = 180^\circ$ and settles close to that location, thus failing in its attempt to switch. This happened because of  the coupled $\theta$-$\phi$ dynamics 
that resulted in a misdirected torque when the magnetization reached the $x$-$y$ plane.
This kind of dynamics has been explained in Ref. \onlinecite{roy11_5}.

\section{\label{sec:conclusions}Conclusions}

We have theoretically investigated  stress-induced switching of multiferroic  nanomagnets in the presence of thermal fluctuations. 
The room-temperature thermal average of the energy dissipation is as small as $\sim$200 $kT$ while the thermal average of the switching delay is  $\sim$0.5 ns with a standard deviation less than 0.1 ns.
This makes strain-switched multiferroic nanomagnets very attractive platforms for implementing non-volatile memory and logic systems because they are minimally dissipative while being 
adequately fast. Our results also show that a certain critical stress is required to switch with $\sim$100\% probability in the presence
of thermal noise. The value of this critical stress increases with decreasing ramp rate until the ramp
rate becomes so slow that $\sim$100\% switching probability becomes unachievable. Thus, a faster 
ramp rate is beneficial. The energy dissipations and switching delays are roughly independent of ramp rate if switching is 
always performed with the 
critical stress.

\vspace*{2mm}

This work was supported by the US National Science Foundation under Nanoelectronics for the year 2020 grant ECCS-1124714 and by the Semiconductor Research Corporation under the Nanoelectronics Research Initiative.


%

\end{document}